\documentstyle[preprint,aps,prl]{revtex}
\begin{document}

\draft

\title{Absolute frequency measurement of the In$^{+}$ clock transition with a mode-locked laser}

\author{J. von Zanthier, Th. Becker, M. Eichenseer, A. Yu. Nevsky, Ch. Schwedes,\\
E. Peik, H. Walther\\
R. Holzwarth, J. Reichert, Th. Udem, T. W. H\"ansch}
\address{Max-Planck-Institut f\"ur Quantenoptik and\\ Sektion Physik der Ludwig-Maximilians-Universit\"at M\"unchen, 85748 Garching, Germany}
\author{P. V. Pokasov, M. N. Skvortsov, S. N. Bagayev}
\address{Institute of Laser Physics, 630090 Novosibirsk, Russia}

\date{\today}

\maketitle

\begin{abstract}

The absolute frequency of the In$^{+}$ $5s^{2 \, 1}S_{0}$ - $5s5p^{\, 3}P_{0}$ clock transition at 237 nm was measured with an accuracy of 1.8 parts in $10^{13}$. Using a phase-coherent frequency chain, we compared the $^{1}S_{0}$ - $^{3}P_{0}$ transition with a methane-stabilized He-Ne laser at 3.39 $\mu$m which was calibrated against an atomic cesium fountain clock. A frequency gap of 37 THz at the fourth harmonic of the He-Ne standard was bridged by a frequency comb generated by a mode-locked femtosecond laser. The frequency of the In$^{+}$ clock transition was found to be $1\;267\;402\;452\;899.92\;(0.23)$ kHz, the accuracy being limited by the uncertainty of the He-Ne laser reference. This represents an improvement in accuracy of more than 2 orders of magnitude on previous measurements of the line and now stands as the most accurate measurement of an optical transition in a single ion.

\end{abstract}

\vspace{1cm}

Improving the accuracy and stability of frequency standards using narrow transitions in atomic systems at optical or higher frequencies as a reference for laser oscillators has been a long-standing objective in metrology \cite{Woods95}. Impressive progress has been achieved in short-term stability of oscillators operating in these spectral regions \cite{Berquist99} as well as in long-term stabilization by directly locking lasers onto narrow atomic transitions \cite{Wine91,Schnatz96,Madej98}. Major advances have recently been reported in the ability to determine absolute optical frequencies using harmonic frequency chains \cite{Schnatz96,Madej99} or frequency chains based on precise measurement of large frequency intervals \cite{Reichert00}.  

Highly accurate and stable optical frequency standards enable precise measurement of fundamental constants \cite{Ude99,Bira99} or investigation of their possible variation in time \cite{Flamm99}. Better frequency standards permit more accurate determination of atomic transitions in spectroscopy and offer the possibility of stringent testing of QED or general relativity. They are also needed in applications such as navigation or very long baseline interferometry \cite{Ramsey95}.

We study the $5s^{2 \, 1}S_{0}$ - $5s5p^{\, 3}P_{0}$ transition at 237 nm in a single trapped In$^{+}$ ion with a view to applying it as an optical frequency standard \cite{Peik94,Peik95,Peik99,Becker00}. This transition with a natural linewidth of only 0.8 Hz \cite{Becker00} has a line quality factor of $Q = \nu / \Delta \nu = 1.6 \times 10^{15}$. The two states participating in the hyperfine-induced $J = 0 \rightarrow J = 0$ transition couple only weakly to perturbing external electric or magnetic fields. In particular, they are insensitive to electric quadrupole shifts caused by the field gradient of the trap. The particle can be kept in the trap almost free of collisions for an infinite time and brought to near-rest by laser cooling. Thus, transit-time broadening and first-order Doppler effects can be eliminated and the second-order Doppler effect can be reduced to negligible values \cite{Peik99,Becker00}. With these characteristics taken into account, a residual uncertainty of  $10^{-18}$ for a frequency standard based on a single stored In$^{+}$ ion is expected \cite{Deh82,Peik94}.

In a recent publication \cite{Zanthier99}, we reported the measurement of the absolute frequency of the In$^{+}$ clock transition using two optical reference frequencies, a methane-stabilized He-Ne laser at 3.39 $\mu$m and a Nd:YAG laser at 1064 nm whose second harmonic was locked to a hyperfine component in molecular iodine. The measurement inaccuracy of 3.3 parts in 10$^{11}$ was limited by the degree of uncertainty to which the iodine reference was known. Here we report on a new precise measurement of the absolute frequency of the $^{1}S_{0}$ - $^{3}P_{0}$ transition using a phase-coherent frequency chain which links the 237 nm radiation (1 267 THz) of the In$^{+}$ clock transition to the He-Ne laser at 3.39 $\mu$m (88 THz) alone. In this case the accuracy of the measurement is only limited by the much smaller uncertainty of the CH$_{4}$/He-Ne reference.

The transportable He-Ne standard was built at the Institute of Laser Physics in Novosibirsk, Russia \cite{HeNe}. It was previously (1996) calibrated for a measurement of the hydrogen 1S - 2S absolute frequency \cite{1s2s_old}. Here we use the result of a more recent calibration \cite{HeNecal} that was obtained together with a new cesium-based measurement of the hydrogen 1S - 2S interval \cite{Niering00}, with $f_{He-Ne}=88\;376\;182\;599\;976\;(10)\;$ Hz. This value deviates from the previous one by 39 Hz (1.6 combined standard deviations), most likely because the operating parameters were not exactly maintained for several years. The new calibration was done 4 months before the measurement of the In$^+$ clock transition. Unlike in the previous calibration, the laser was not moved between its calibration and the measurement.

The experimental setup of the frequency chain is illustrated in Fig.
\ref{scheme}. A NaCl:OH$^-$ color center laser at 1.70 $\mu$m is phase-locked to the second harmonic of the methane-stabilized He-Ne laser at
3.39 $\mu$m. A 848 nm laser diode is then locked to the second harmonic
of the color center laser. This is accomplished by first stabilizing it
to a selected mode of the frequency comb of a Kerr-lens mode-locked
Ti:sapphire femtosecond laser (coherent model Mira 900), frequency-broadened by a standard single-mode silica fiber (Newport FS-F), and then
controlling the position of the comb in frequency space \cite{Reichert00}.
A radio-frequency divider, dividing by 128, helps to overcome the limited servo bandwidth that controls the comb position. Recently, we 
demonstrated that this frequency comb meets the exceptional requirements
of high-precision spectroscopy \cite{Holzwarth00}.
The 76 MHz pulse repetition rate, which sets the mode separation, is phase-locked \cite{Reichert99a} to a commercial cesium clock (Hewlett-Packard
model 5071 A) which is constantly calibrated using the time disseminated
by the Global Positioning System. Each mode of the comb is therefore known
with the same fractional precision as the He-Ne standard, i.e. within
1.1 parts in $10^{13}$ \cite{HeNecal}. A diode laser at 946 nm is
phase-locked to another selected mode of the comb, positioned $n=482\;285$
modes or 37 THz to lower frequencies from the initial mode at 848 nm.
The mode number $n$ is determined from our previous measurement
\cite{Zanthier99}. The beat frequency with the 946 nm Nd:YAG laser, whose
4th harmonic excites the In$^+$ clock transition, is counted with a
commercial radio-frequency counter.

For the absolute frequency measurement, the in-lock beat signals of the
chain are continuously monitored with additional
counters for possible lost cycles \cite{Reichert00}. Points that are off by more than 0.5
Hz from the phase-locked beat signal are discarded. With the frequency
chain in lock, the unknown $^{115}$In$^{+}$ $^{1}S_{0}$ - $^{3}P_{0}$ clock
transition frequency $f_{In^{+}}$ at 237 nm is related to the known
frequency of the He-Ne standard $f_{He-Ne}$ through

\begin{equation} \label{freq1}
f_{In^{+}} = 16 \cdot f_{He-Ne} - 4 \cdot ( f_{B} + n \cdot f_{r} ) - f_{LO}.
\end{equation}

Here $f_B$ is the frequency of the beat signal at 946 nm detected by a
photodiode and recorded by a radio-frequency counter in a 1 sec interval,
$f_{r}$ denotes the repetition rate of the mode-locked femtosecond laser,
$n$ is the number of modes separating the two selected modes of the comb, and
$f_{LO}$ = 1 632 MHz contains all contributions from the local oscillator
frequencies employed for the phase-locks.

Within several days of measurement we recorded 6 214 excitations of the $^{3}P_{0}$ level by scanning the 946 nm Nd:YAG clock laser over the In$^{+}$ $^{1}S_{0}$ - $^{3}P_{0}$ resonance. Excitation of the $^{3}P_{0}$ state is detected by optical-optical double resonance \cite{Deh82}: after applying the clock laser for 15 ms the laser exciting the $^{1}S_{0}$ - $^{3}P_{1}$ cooling transition is turned on for 40 ms to probe the population of the ground state. If no fluorescence photons are counted on the $^{1}S_{0}$ - $^{3}P_{1}$ transition, the cooling laser is kept on for up to ten further 40 ms intervals to wait for the decay of the metastable state and an excitation event is recorded. The frequency of the clock laser is typically changed in steps of 80 Hz and 16 excitation attempts are made at each frequency. Scanning of the Nd:YAG master laser is synchronized with detection of the beat signal $f_B$ at 946 nm and the counters used for detection of lost cycles. All counters as well as the AOM driving synthesizers are referenced to a local cesium atomic clock.

Figure \ref{inres}a shows the $^{3}P_{0}$ excitation probability as a function of the beat frequency $f_B$ for a typical measurement session, collecting 674 quantum jumps to the $^{3}P_{0}$ state. During the measurement session 21 excitation spectra were recorded where the investigated scanning range varied slightly from spectrum to spectrum. The width of the distribution is due to short-term and mid-term frequency instabilities of the Nd:YAG laser, the methane standard, and the frequency chain. A weighted fit to a Gaussian function is used to determine the line center. By averaging the beat frequency at line center for the eleven measurement sessions performed we obtain $f_B = 49\;174\;925\;(42)\;$ Hz (see Fig. \ref{inres}b). From this value we determine the absolute frequency $f_{In^{+}}$ of the $^{115}$In$^{+}$ $^{1}S_{0}$ - $^{3}P_{0}$ clock transition according to Eq. (\ref{freq1}):

$$
f_{In^{+}} = 1\;267\;402\;452\;899.\;92\;(0.23)\;\textrm{kHz}.
$$

The uncertainty of 1.8 parts in 10$^{13}$ is derived from a quadratic addition of the uncertainty of the measurement (4 $\Delta f_B$) and the uncertainty in the He-Ne standard calibration (16 $\Delta f_{He-Ne}$). Both contributions are due to the limited reproducibility of the He-Ne standard.

Systematic frequency shifts of the clock transition are negligible at the present level
of accuracy. The magnetic field dependence is -636 $\pm$ 27 Hz/G
for the $\mid F = 9/2 ; m_{F} = 9/2 > \, \rightarrow \, \mid F =
9/2; m_{F} = 7/2 >$ component that we excite \cite{Becker00}, where we adjust the 
magnetic field to zero with a precision of a few times 10 mG. Other systematic frequency
shifts such as the quadratic Stark or second-order Doppler shift
are orders of magnitude smaller than the Zeeman shift at the temperatures to which
the ion is cooled in our trap (T $\sim$ 150 $\mu$K) \cite{Becker00}.

The new result is well within the error bars of our previous measurement, where we obtained a value of $1\;267\;402\;452\;914\;(41)$ kHz  for the clock transition \cite{Zanthier99}. Compared with that value, the new measurement represents an improvement in accuracy of more than 2 orders of magnitude. It now stands as the most accurate measurement of an optical transition frequency in a single ion. For the future it is planned to use a femtosecond laser frequency comb to compare the indium clock transition directly with a cesium clock. In this case the cesium reference could be provided by a transportable fountain clock, as was done in a recent measurement of the 1S  - 2S transition in hydrogen \cite{Niering00}. The stability and reproducibility of the $^{1}S_{0}$ - $^{3}P_{0}$ clock transition makes indium an attractive candidate for such ultra-precise frequency measurement. 

A further interesting feature of our frequency chain is the fact that it can be used to compare two narrow optical transition frequencies with each other, e.g. the clock transition in indium with the hydrogen 1S - 2S transition. In particular, this may allow investigation of possible variations of fundamental constants in time as recently proposed in \cite{Flamm99}.

The collaboration with ILP, Novosibirsk, was supported by the Volkswagen Foundation within Project No. I/72 607.

\begin{figure}[htbp]
\caption{Set-up of the frequency chain used to measure the absolute frequency of the In$^{+}$ clock transition. The chain links the 237 nm radiation of the clock transition (1 267 THz) to a methane-stabilized He-Ne laser at 3.39 $\mu$m (88 THz). A frequency gap of 37 THz at 848 nm is bridged with the help of a frequency comb created by a Kerr-lens mode-locked femtosecond laser.}
\label{scheme}
\end{figure}

\vspace{10mm}

\begin{figure}[htbp]
\caption{a) Excitation probability of the $^{3}P_{0}$ state as a function of the beat frequency $f_B$ at 946 nm for a typical measurement session. 674 quantum jumps to the $^{3}P_{0}$ level were recorded in this session. Error bars correspond to the reciprocal number of $^{3}P_{0}$ excitation trials in a 30 Hz frequency interval. They are used as weights for the fit to a Gaussian curve. b) Beat frequency $f_B$ at line center for the eleven measurement sessions performed. The mean value is $f_B = 49\;174\;925\;(42)\;$ Hz. Error bars are given by the uncertainty of the He-Ne standard.}
\label{inres}
\end{figure}

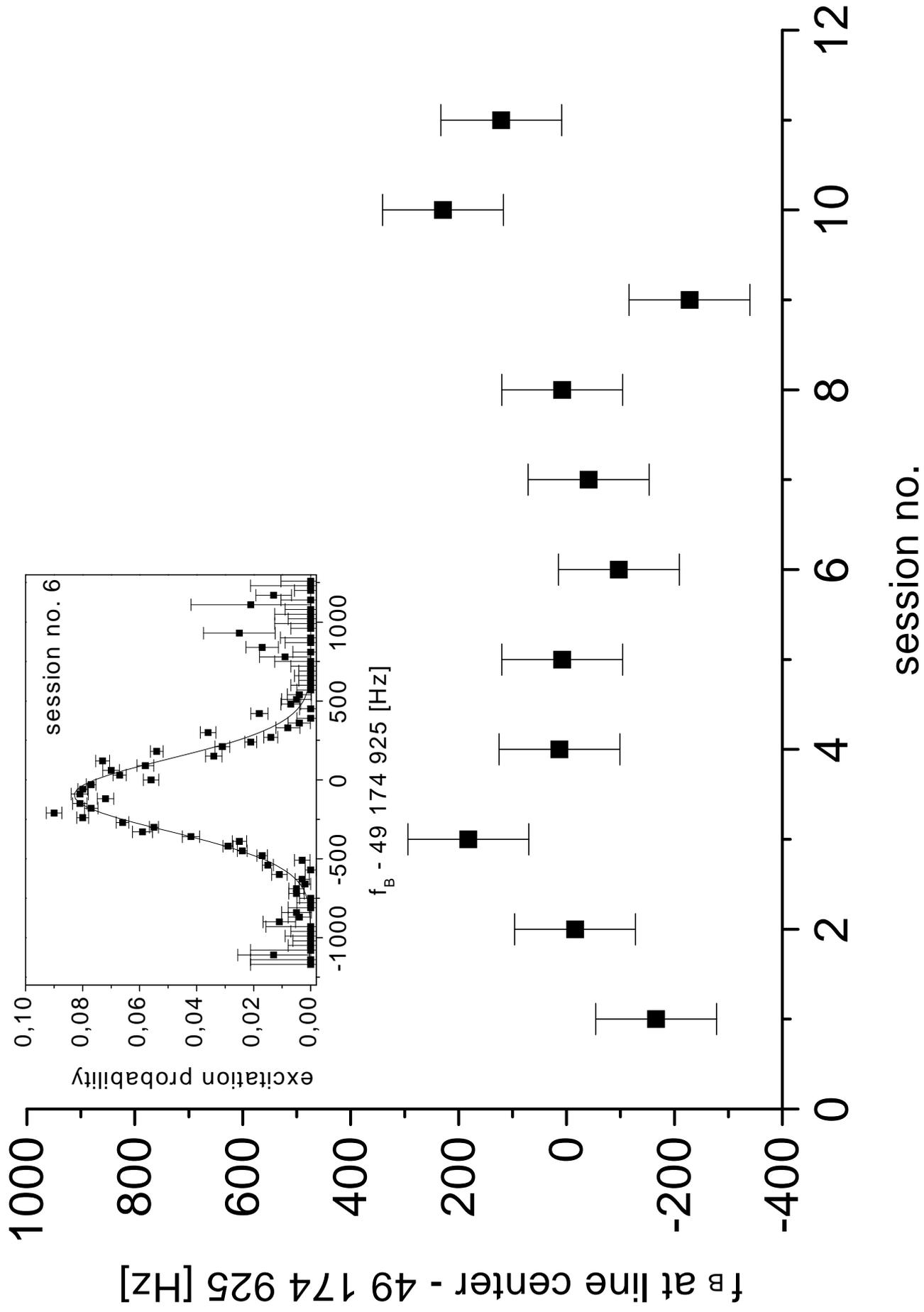
\end{document}